\documentstyle[twocolumn]{IEEEtran}

\input epsf

\begin{document}

\title{All-dielectric  one-dimensional periodic structures for total omnidirectional reflection and partial spontaneous emission control}

\author{D.~N. Chigrin\thanks{D.~N. Chigrin is with the Universit{\" a}t GH Essen, Fachbereich Physik, D-45117 Essen, Germany. E-mail: dmitry.chigrin@physics.org. Author to whom correspondence should be addressed.}, A.~V. Lavrinenko\thanks{A.~V. Lavrinenko is with the Belarusian State University, Department of  Physics, Fr. Skarina Ave. 4, 220080 Minsk, Belarus. E-mail: lavrinenko@phys.bsu.unibel.by.}, D.~A. Yarotsky, and S.~V. Gaponenko\thanks{D.~A. Yarotsky, and S.~V. Gaponenko are with the Institute of Molecular and Atomic Physics, National Academy of Sciences, 220072 Minsk, Belarus. E-mail: gaponen@imaph.bas-net.by.}}

\markboth{IEEE Journal of Lightwave Technology}
{Chigrin at el.: All-dielectric  one-dimensional periodic structure...}

\maketitle

\begin{abstract}
A remarkable property of one-dimensional all-dielectric periodic structures has recently  been reported, namely a one-dimensional lattice can totally reflect electromagnetic wave of any polarization at all angles within a prescribed frequency region. Unlike their  metallic counterpart, such all-dielectric omnidirectional mirrors are nearly free of loss at optical frequencies. Here we discuss the physics, design criteria and applications of the thin-film all-dielectric omnidirectional mirror. The experimental demonstration of the mirror is presented at optical frequencies.
\end{abstract}


\section{Introduction}
\PARstart{I}{t} is well known that spontaneous emission can be strongly modified by changing the environment near an excited atom. A microcavity with perfectly reflecting walls can considerably inhibit or enhance spontaneous emission of atoms placed inside it~\cite{cavityQED1,cavityQED2}.

Metallic mirrors being a good reflector for any angle of incidence fulfill the requirements to be inside walls of microcavities. At optical frequencies there is however a problem with them: they display notable dissipative losses.

Photonic crystals were originally proposed by Yablo\-no\-vitch~\cite{Basic1} to fix this problem. Photonic crystals are periodically microstructured dielectric materials which can exhibit the frequency bands that are completely free of electromagnetic states. The forbidden band is usually referred to as a {\em full} three-dimensional (3D) photonic band gap. Being made from positive-dielectric-constant materials photonic crystal can be almost free of dissipative losses at any prescribed frequency. In the limit of a thick sample photonic crystal behaves as an omnidirectional high reflector.
 
Since the first works~\cite{Basic1,Basic2}, the concept of photonic crystal has been attracting close attention of the scientific community and a lot of applications have been proposed (see ~\cite{cavityQED2,PBGs1,PBGs2,PBGs3,PBGs4,Joannopoulos} and refs. therein).

However, until now the possibility to design an all-dielectric microstructure displaying total omnidirectional reflection has mainly been associated with 3D periodic materials.  In spite of current success in the microstucturing of 3D photonic crystals~\cite{OpticalPBG1,OpticalPBG2,OpticalPBG3,OpticalPBG4}, there is still a serious technological problem to fabricate a periodic structure of arbitrary wavelength-scale period. The investigations of low dimensional periodic media have been attracting considerable interest.

It has recently been recognized, that 2D~\cite{Russell2D} and 1D~\cite{PRL,Winn,Fink,APA99,Russell99,Dowling} periodic structures can display some features of full 3D photonic band gap, namely, display total omnidirectional reflection of arbitrary polarized wave within certain frequency region.

A one-dimensional photonic crystal is nothing other than the well-known dielectric Bragg mirror consisting of alternating layers with low and high indices of refraction. In contrary to 3D microstructures, 1D dielectric lattices are unique in that modern technology is currently able to produce the needed wavelength-scale period.

Depending on the chosen geometry and frequency region, a lot of applications are possible. The planar geometry, for example, can be used to improve the properties of vertical-cavity surface-emitting lasers and microwave antenna, to design transmission and energy saving filters. By rolling into hollow fibers, the mirror can be used as inside walls of high-finesse waveguides and microcavities.

The paper is organized as follows. Section II outlines the origin of the total omnidirectional reflection displayed by 1D photonic crystals. Optimization criteria of the mirror design are reported. The possibility of partial spontaneous emission control is discussed. In Section III,  the experimental demonstration of the mirror is presented at optical frequencies.

\section{Theoretical results}

\subsection{Total omnidirectional reflection}

\begin{figure}[t]
\noindent \epsfbox{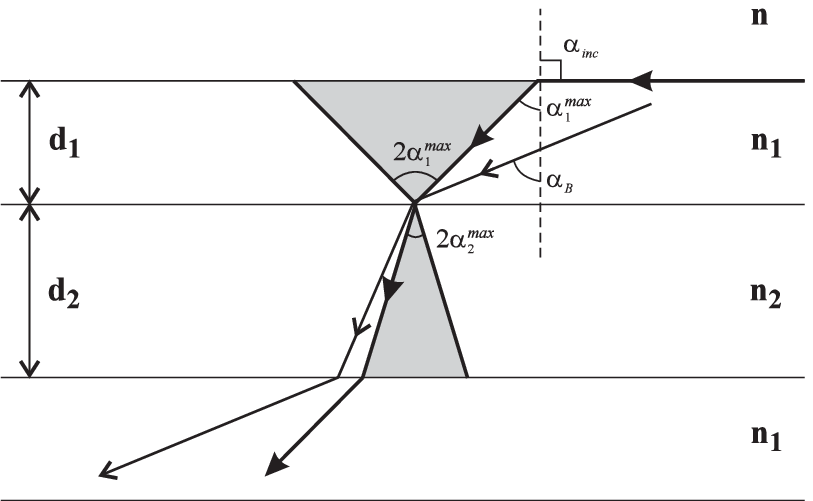}
\caption{Schematic representation of a dielectric multilayer structure of alternating layers of low, $n_1$, and high, $n_2$, indices of refraction. $d_1$ and $d_2$ are the  thicknesses of the layers. $n$ is a reflective index of an ambient medium. The light rays refracting and propagating through a stack are presented.  The full domain of incident angles $\alpha_{inc} \in [-\pi /2, \pi /2]$ is mapped into the internal cone of half-angle $\alpha^{max}_1=\arcsin n/n_1$ (the light gray area). For sufficiently large index ration $\delta n_0=n_1/n$ the internal cone's half-angle $\alpha^{max}_1$ may be smaller than the Brewster angle  $\alpha_B$. {\em An externally incident wave can never couple to the Brewster window}.}
\label{Rays}
\end{figure}

Consider an infinite periodic stack of alternating layers of low, $n_1$, and high, $n_2$, indices of refraction and the thicknesses $d_1$ and $d_2$, respectively (Fig.~\ref{Rays}). The period is $\Lambda=d_1 + d_2$. The periodicity of the structure leads to the Bloch wave solutions of the Maxwell equations. The Bloch wave number $K$ may be obtain from the dispersion relation (see e.g.~\cite{Yeh,Russell95})
\begin{equation}
K({\bf k}_\perp, \omega) {\em \Lambda}=\arccos \left( \frac{1}{2} (A+D) \right) \label{dispersion}
\end{equation}
where ${\bf k}_\perp$ is the tangential component of the Bloch wave vector, $\omega$ is the frequency. A particular form of the quantities $A$ and $D$ may be found elsewhere~\cite{Yeh,Russell95}. Due to the planar geometry of the problem, the separation of the electromagnetic field into TE (transverse electric) and TM (transverse magnetic) polarization states is possible, where the electric or magnetic field vector, respectively, is parallel to the layers interfaces. This splits the problem into the two independent ones for TE and TM polarizations, respectively.

\begin{figure}[b]
\noindent \epsfxsize=8.6cm \epsfbox{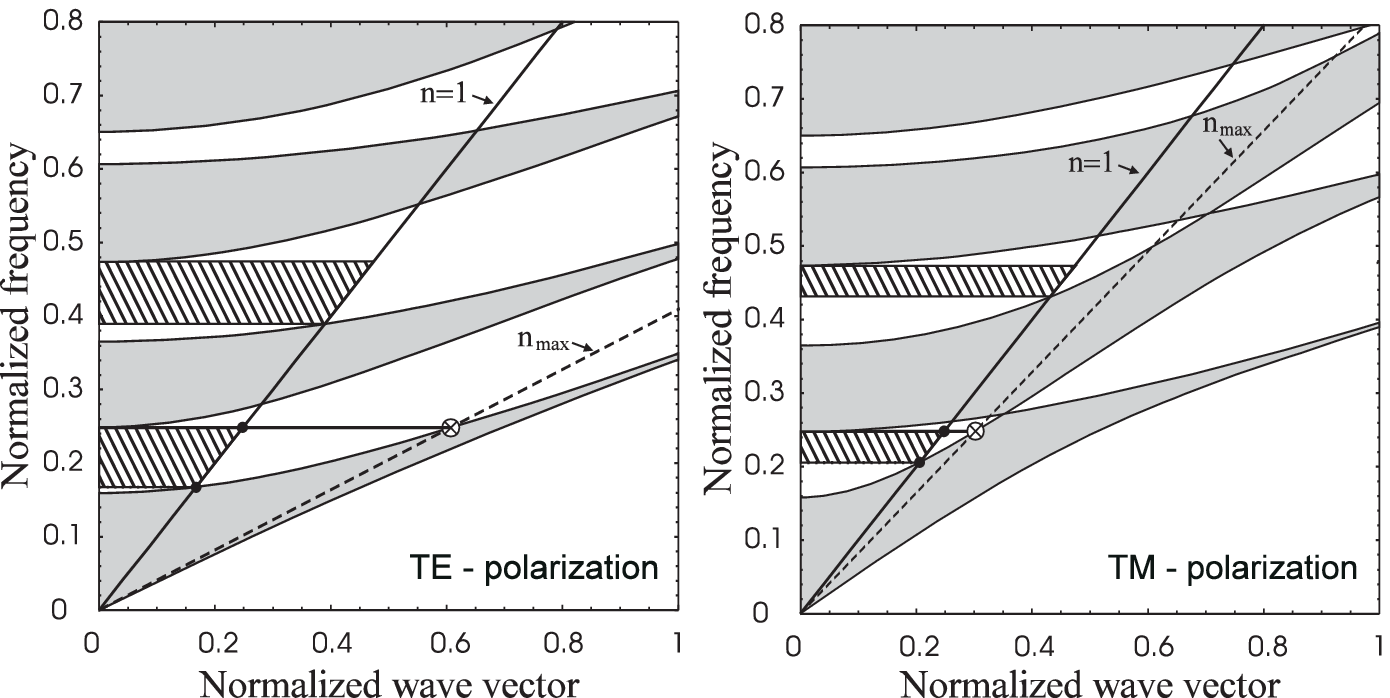}
\caption{Photonic band structure of a typical 1D photonic crystal for TE (left panel) and TM (rigth panel) polarizations. The stack is assumed of refractive indices $n_1=1.4$, $n_2=3.4$ and filling fraction $\eta=d_2/\Lambda=0.5$. The frequency and the tangential component of the wave vector are defined to be normalized as $\omega \Lambda /2\pi c$  and $\left|{\bf k}_\perp\right|\Lambda/2\pi$, respectively. The gray areas correspond to the propagating states, whereas  white areas contain the evanescent states only. The shaded areas correspond to omnidirectional reflection bands. The solid and dashed lines are an ambient-medium light-lines.}
\label{PBGk}
\end{figure}

Photonic band structure of an infinite system of layers is depicted in the figure~\ref{PBGk}. The refractive indices, $n_1=1.4$ and $n_2=3.4$, are chosen close to ones of SiO$_2$ and Si in the near IR region, where these materials are essentially transparent. The band structure has been calculated using the dispersion equation~(\ref{dispersion}). The left panel is for TE polarization, and the right one for TM. An infinite periodic structure can support both propagating and evanescent Bloch waves, depending on real or imaginary Bloch wave numbers are. In the figure~\ref{PBGk}, gray areas correspond to the propagating states, whereas  white areas contain the evanescent states only and are usually referred to as photonic band gaps.

\begin{figure}[t]
\noindent \epsfbox{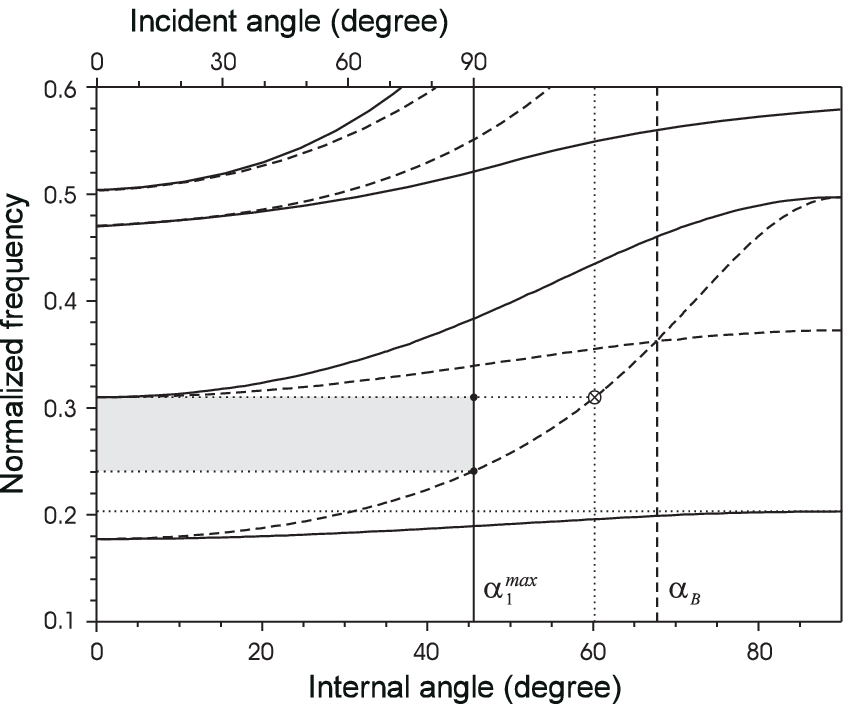}
\caption{Photonic band structure in terms of normalized frequency and the internal angle in the low index layer, $n_1$. The stack is assumed of refractive indices $n_1=1.4$, $n_2=3.4$ and filling fraction $\eta=0.32$. The solid (dashed) curves are for TE- (TM-) polarization bands. The internal angle, for which the TM forbidden gap is closed up, is depicted by the vertical dotted line. The solid vertical line corresponds to the internal cone's half-angle $\alpha^{max}_1$ in the case of the ambient medium with refractive index $n=1$. The Brewster angle $\alpha_B$ is depicted by the dashed vertical lines. Light gray area are for the omnidirectional total reflection band. }
\label{PBGa}
\end{figure}

As only the normal component of the wave vector is involved in the band gaps formation for oblique propagation, the band gaps shift towards the higher frequencies with the tangential component of the wave vector  (Fig.~\ref{PBGk}). The common feature of the band structure of 1D lattices is that the forbidden gaps are always closed up (e. g. the crossed circles in figure~\ref{PBGk}). Another feature of the band structure is that the TM forbidden gaps shrink to zero onto the Brewster light-line, where $\omega=c \left|{\bf k}_\perp\right|/n_1\sin{\alpha_B}$ (Fig.~\ref{PBGk}), $\alpha_B=\arctan n_2/n_1$ is the Brewster angle. The TM polarized wave propagates without any reflection from $n_1$ to $n_2$ layer, and from $n_2$ to $n_1$ layer, at the Brewster angle $\alpha_B$.

Suppose that the plane electromagnetic wave illuminates the boundary of a semi-infinite periodic stack under the angle, $\alpha_{inc}$, from semi-infinite homogeneous medium of refractive index, $n$ (Fig.~\ref{Rays}). While the frequency and the wave vector of an incident wave are fitted into the forbidden gaps of the photonic crystal\footnote{It is important to note that infinite and semi-infinite photonic crystals have the same band structure~\cite{Zolla}, the only difference is the existence of surface modes in the case of semi-infinite stack.}, an incident wave undergoes total reflection. The band gaps of the crystal lead to the total reflection bands in the spectra, which are very sensitive to the incident angle. Two questions arise: (i) whether it is possible to avoid the coupling of the incident wave to the Brewster window, where the reflection coefficient at the interface of low, $n_1$, and high, $n_2$, index layers is identically zero; and (ii) whether the reflection bands (photonic band gaps) can be wide enough to be open for all incident angles.

When electromagnetic wave illuminates the boundary of the semi-infinite crystal, the possible values of the internal angles are restricted by the Snell's law. The full domain of incident angles $[-\pi /2, \pi /2]$ is mapped into the internal cone of half-angle $\alpha^{max}_1=\arcsin n/n_1$ (the light gray area in the figure~\ref{Rays}). The larger are refractive indices of the layers with respect to the medium outside the crystal, the narrower is a cone of internal angles. For sufficiently large index ratio $\delta n_0=n_1/n$ the internal cone's half-angle $\alpha_{1}^{max}$ can be smaller than the Brewster angle $\alpha_{B}$. An externally incident wave will never couple to the Brewster window. This answers the first question.

\begin{figure}[t]
\noindent \epsfbox{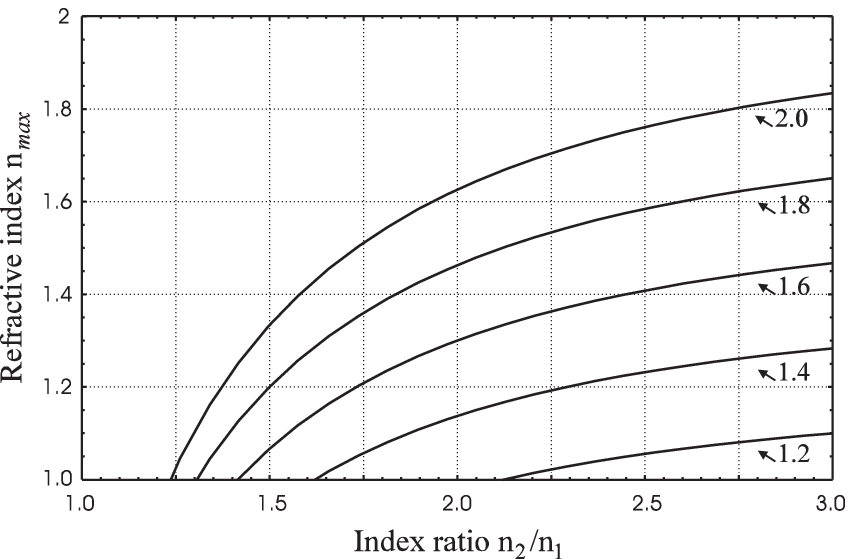}
\caption{Maximum refractive index of the ambient medium, for which an omnidirectional reflection band is closed up, as a function of the index ratio $\delta n=n_2/n_1$ for different values of refractive index of the low index layer, $n_1$, and fixed filling fraction $\eta=0.5$. Numbers at the curves indicate the refractive index of the low index layer, $n_1$.}
\label{MaxN}
\end{figure}

\begin{figure}[h]
\noindent \epsfbox{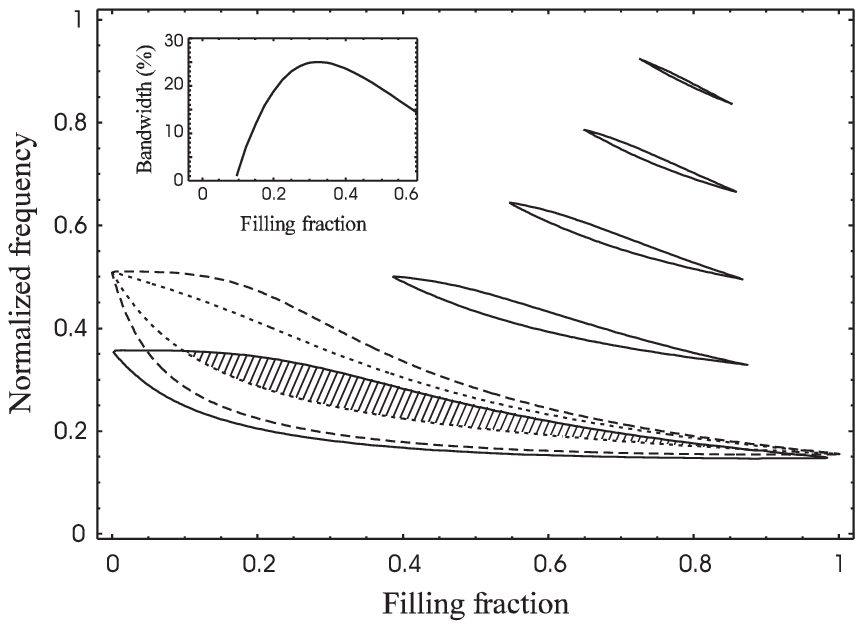}
\caption{Overall band gaps leading to the omnidirectional total reflection bands. For the first overall band gap, gaps corresponding to the normal and grazing incidence are presented. The shaded area are for the first omnidirectional reflection band. The solid curves is for normal incidence. The dashed (dotted) curves is for grazing incidence for TE (TM) polarization. The inset shows relative bandwidth versus filling fraction. Here $n=1$, $n_1=1.4$ and $n_2=3.4$.}
\label{Eta}
\end{figure}

To answer the second question, consider the reduced region of k-space associated with the ambient medium (Fig.~\ref{PBGk}). For an incident wave with the wave vector $\left|{\bf k}\right|=n\omega/c$, the tangential component of the wave vector remains constant throughout the crystal and equals to $\left|{\bf k}_\perp\right|=(n\omega/c)a\sin{\alpha_{inc}}$. Here, $\alpha_{inc}$ is incident angle, $\omega$ is the frequency and $c$ is the speed of light in vacuum. The wave coming from the outside can only excite the states lying above the ambient-medium light-line (solid line in the figure~\ref{PBGk} corresponds to the ambient medium with refractive index $n=1$). To have an omnidirectional reflection the forbidden gap should be open within this reduced region of k-space. A sufficiently large index ratio $\delta n = n_2/n_1$ can make a trick, leading to the wide band gap opened for all incident angles.

For the structure presented (Fig.~\ref{PBGk}) and air as an ambient medium, $n=1$, both index ratio $\delta n= n_2/n_1 \approx 2.4$ and $\delta n_0=n_1/n=1.4$ are sufficiently large, so the first two overall band gaps are open for all external angles of incidence. No propagating mode are allowed in the stack for any propagating mode in the ambient medium within the gaps for both TE and TM polarizations (shaded areas in the figure~\ref{PBGk}). {\em The total omnidirectional} reflection arises. A band of surface modes lies below the ambient-medium light-line~\cite{Joannopoulos} and thus surface modes do not affect the external reflectivity.

It is instructive also to represent the band structure in terms of internal angles. An internal angle parametrizes the tangential component of the wave vector as $\left|{\bf k}_\perp\right|=(n_i\omega/c)\sin{\alpha_{int}}$, where $n_i$ is the reflective index of the layer,  $\alpha_{int}$ is the internal angle in the $n_i$ layer. In the figure~\ref{PBGa} the band structure is redrawn in terms of internal angles in the low index layer. One can see that the overall forbidden gap is opened for all external incident angles for both fundamental polarizations forming {\em an omnidirectional total reflection band} (gray area in the figure~\ref{PBGa}). 

\begin{figure}[t]
\noindent \epsfbox{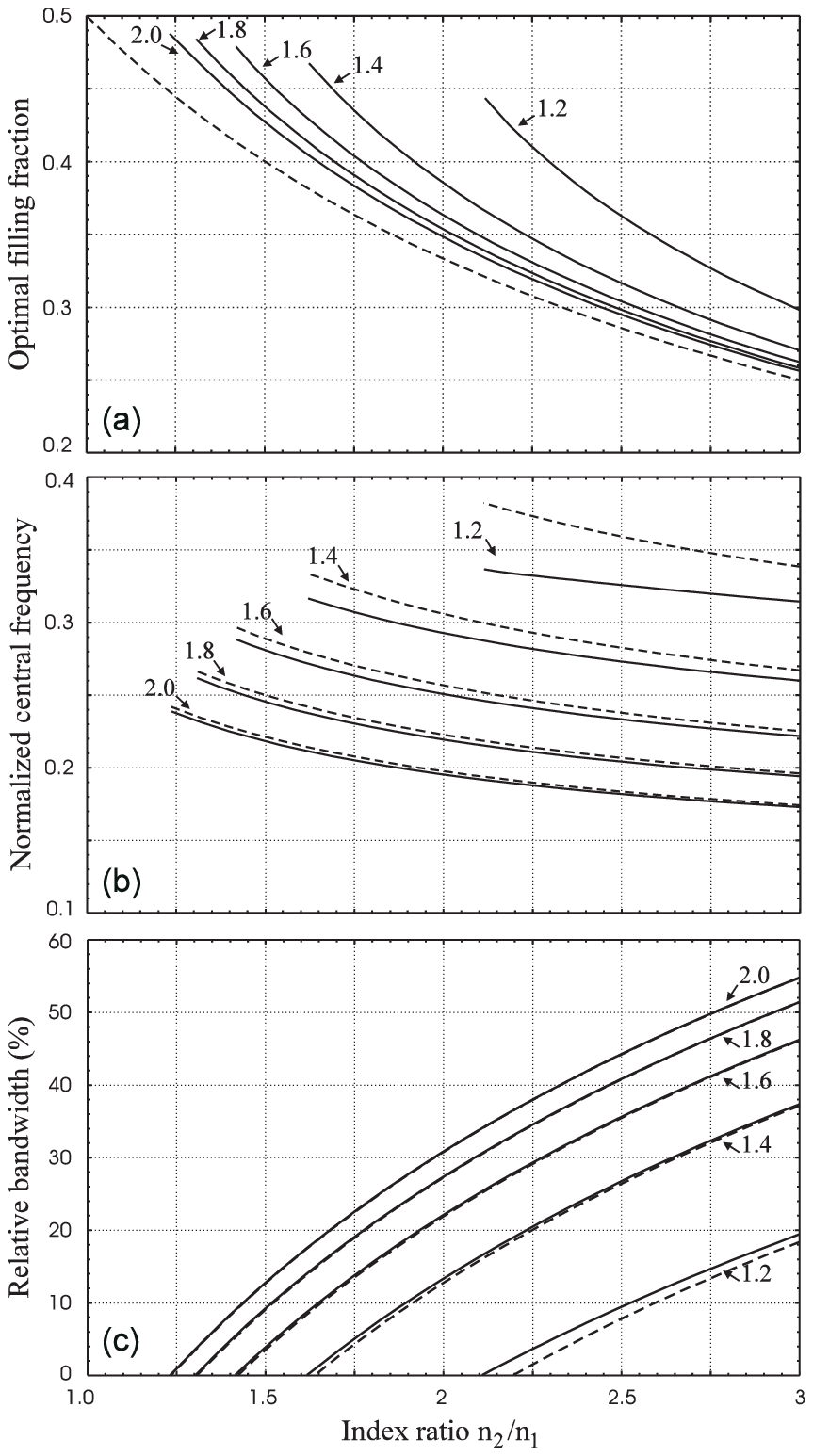}
\caption{(a) Optimal filling factor $\eta_{opt}$ versus the index ratio $\delta n$ for different values of index ratio $\delta n_0$. (b) Central frequency $\omega_0$ versus the  index ratio $\delta n$ for different values of index ratio $\delta n_0$ for optimal filling factor  $\eta_{opt}$. (c) Relative bandwidth $\Delta \omega/\omega_0$ versus the index ratio $\delta n$ for different values of index ratio $\delta n_0$ for optimal filling factor  $\eta_{opt}$. Numbers at the curves indicate the index ratio $\delta n_0$. The dashed curves correspond to the quarter-wave stack.}
\label{Optimal}
\end{figure}

\subsection{Optimization criteria}

To design an omnidirectional mirror, one needs to have total reflection for all incident angles and all polarization states. The TM photonic band gap is narrower than TE one and so defines the bandwidth of an omnidirectional reflection band. The upper edge of the reflection band corresponds to the upper edge of the forbidden gap at normal incidence. The lower edge is defined by the intersection of the ambient-medium light-line with the upper edge of the corresponding TM band (Fig.~\ref{PBGk}). For given parameters of the periodic structure, the refractive index of an ambient medium may be used to control the bandwidth. By increasing the refractive index of an ambient medium from $n=1$ to some $n=n_{max}$ the bandwidth decreases till an omnidirectional reflection band is closed up. In figure~\ref{MaxN} the maximum refractive index of the ambient medium $n_{max}$, for which the first omnidirectional reflection band is closed up,  is presented as a function of the index ratio $\delta n=n_2/n_1$ for various values of the refractive index of the low index layer $n_1$ and the fixed filling fraction $\eta=d_2/\Lambda=0.5$.

The filling fraction $\eta=d_2/\Lambda$ optimizes the relative reflection bandwidth $\Delta \omega/\omega_0$ of an omnidirectional reflection band with respect to the given refractive indices of the layers constituting the 1D photonic crystal and the index of an ambient medium. Here $\Delta \omega$ is the width of the omnidirectional reflection band and $\omega_0$ is the central frequency. In figure~\ref{Eta} an overall band gaps leading to the omnidirectional total reflection bands are presented versus the filling fraction. For the first overall band gap, gaps corresponding to the normal and grazing  incidence are presented. The solid curves are for normal incidence. The dashed (dotted) curves are for grazing incidence for TE (TM) polarization. The omnidirectional reflection band, which  is due to the overlap of the gaps corresponding to the normal and grazing incidence, is depicted as the shaded area. The omnidirectional reflection bands of the higher order are opened. The inset (Fig.~\ref{Eta}) shows relative bandwidth of the first total reflection band versus filling fraction. There is a clear optimum filling fraction $\eta_{opt}$ leading to the maximum of the relative bandwidth.

We further present the set of contour plots (Fig.~\ref{Optimal}) which provides the full information about the first omnidirectional total reflection band for given parameters of the system. An optimal filling fraction and corresponding central frequency are shown in figures~\ref{Optimal}~(a) and~\ref{Optimal}~(b), respectively, as a functions of the index ratio $\delta n$ for different values of index ratio $\delta n_0$. The dashed curve in figure~\ref{Optimal}~(a) corresponds to the filling factor of a quarter-wave stack, which is $\eta_{\lambda /4}=1/(1+\delta n)$. Within a given parameter range a quarter-wave stack is not an optimal configuration to reach a maximum relative bandwidth for the omnidirectional reflection band, however it gives the relative bandwidth which is usually few percent smaller than optimal one [Fig.~\ref{Optimal}~(c)]. In figure~\ref{Optimal}~(c) the optimal relative bandwidth is depicted as a function of the index ratio $\delta n$ for different values of index ratio $\delta n_0$. A wide omnidirectional total reflection band exists for reasonable values of both $\delta n$ and $\delta n_0$. To obtain an omnidirectional band with bandwidth larger than 5\% 
the index ratios should be larger than 1.5 ($\delta n>1.5, \delta n_0>1.5$).  A decrease in one of the index ratios is partially compensated by an increase in the other one.

For the SiO$_2$/Si ($n_1=1.4$, $n_2=3.4$) structure in air ($n=1.0$) the omnidirectional reflection band is centered at the normalized frequency $\omega \Lambda /2\pi c=0.275$ with the optimal filling fraction $\eta_{opt}=0.324$. The relative bandwidth is about 25\%. 
To obtain an omnidirectional reflection centered at the radiation wavelength $\lambda=1.5~\mu m$, one needs a period of the structure of about $0.412~\mu m$. 

\begin{figure}[t]
\noindent \epsfbox{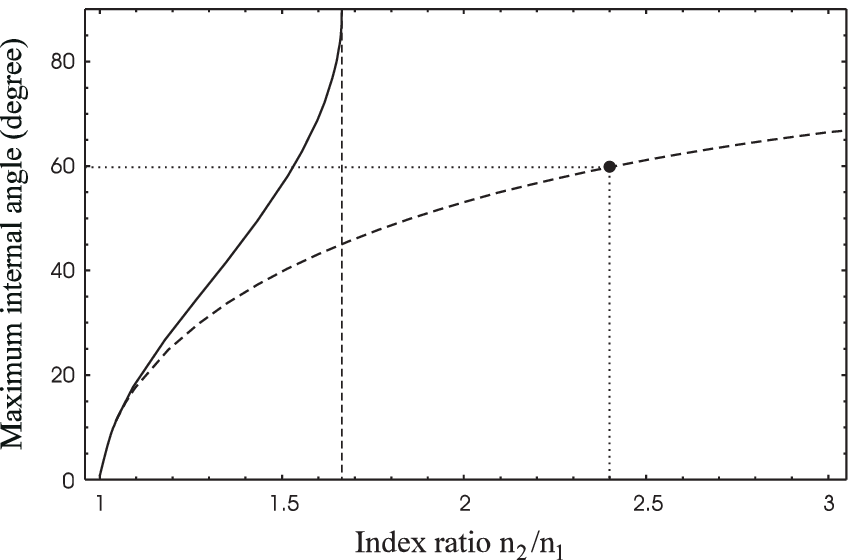}
\caption{An internal angle in low index layer, for which an overall forbidden gap is closed up, as a function of the index ratio $\delta n$. The filling fraction is $\eta=0.32$. The solid curve is for TE polarization, the dashed one is for TM polarization. Black dot corresponds to the stack parameters as in the figure~\ref{PBGa}: $n_1=1.4$, $n_2=3.4$. For all index ratios $\delta n$ which are to the right of the dashed vertical line, TE forbidden gap is open for all internal angles.}
\label{InternalGap}
\end{figure} 

\begin{figure}[t]
\noindent \epsfbox{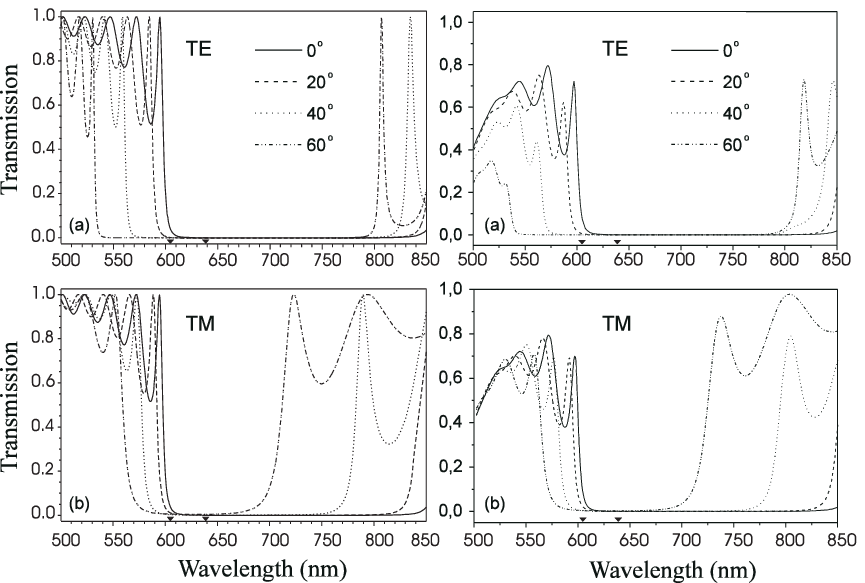}
\caption{Calculated transmission spectra (left panel) and measured transmission spectra (right panel) of Na$_3$AlF$_6$/ZnSe 19-layer structure for TE-polarized (a) and TM-polarized (b) light at different angles of incidence ($0^\circ$--solid line, $20^\circ$--dashed line, $40^\circ$--dotted line, $60^\circ$--dash-dotted line). The triangles indicate the calculated  edges of the omnidirectional reflection band.}
\label{Transmission}
\end{figure} 

\begin{figure}[h]
\noindent \epsfbox{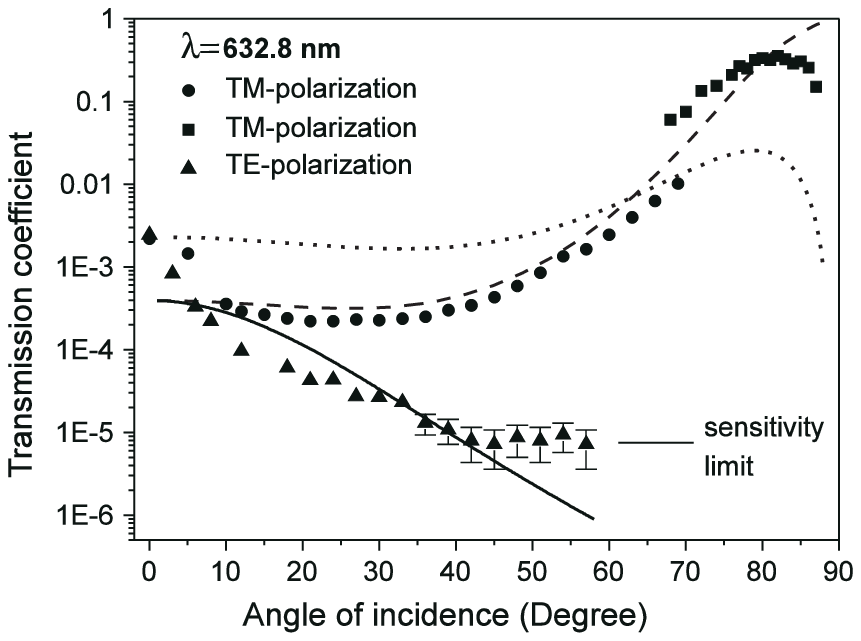}
\caption{Dependence of transmission coefficient of Na$_3$AlF$_6$/ZnSe structure on angles of incidence at wavelength of 632.8~nm (HeNe laser) for two polarizations of incident light. For TM-polarization circles are directly measured transmission coefficients, squares mark transmission coefficients calculated from reflection measurement data. For TE-polarization (upper triangles) the signal at angles more than $60^\circ$ is out of the sensitivity range of the experimental set-up. The solid (dashed) curve is the theoretically calculated transmission coefficient for TE- (TM-) polarized light. The dotted curve corresponds to the transmission coefficient of the structure terminated at both ends with low index layer, Na$_3$AlF$_6$, for TM-polarized light.}
\label{HeNe}
\end{figure}

\subsection{Perfect mirror or more?}

To suppress substantially the spontaneous emission of radiating species placed inside the layer of a periodic stack, the layer must be free of all unbound and bound modes. For the structure presented (Fig.~\ref{PBGa}) an overall forbidden gap is open up to the grazing angles in the case of TE polarized radiation, while an overall forbidden gap is closed up at near $60^\circ$ of an internal angle for TM polarization. The internal angle in the low index layer, for which an overall forbidden gap is closed up, is presented in the figure~\ref{InternalGap} as a function of the index ratio $\delta n$. Black dot corresponds to the stack parameters as in the figure~\ref{PBGa}: $n_1=1.4$, $n_2=3.4$. For all index ratios $\delta n$ which are to the right of the dashed vertical line, an overall TE forbidden gap is open for all internal angles. For such design parameters, the low index layer of the structure can be completely free of TE polarized propagating modes. To ensure the layer is free of all modes, bound modes, which are essentially guided modes, should be absent as well. However, there be no modes within the low index layer which can couple to the high-index-layer modes, including guided modes~\cite{Yeh}.  The radiation of emitting species embedded in low index layer may be inhibited over about $\pm 60^\circ$ aperture of internal angles (Fig.~\ref{PBGa}).

Prospects for partial spontaneous emission control using this kind of  1D periodic structures are discussed in~\cite{Russell99}: introducing a low index defect layer Russell et. al. resolve the Brewster window problem, namely design a defect layer which may be free of all unbound and bound modes. Properly designed defect layer may provide the emission control over a solid angle $4 \pi$ steradian.

Another way to resolve the Brewster window problem is associated with anisotropic periodic medium~\cite{ChigrinLavrinenko}. Forbidden band gaps of such a periodic structure may do not shrink to zero anywhere, so the low index layers of the structure may be free of all electromagnetic modes.

\section{Experimental results}

We have chosen to check theoretical predictions at optical wavelengths. A lattice consisting of 19 layers of Na$_3$AlF$_6$ and ZnSe ($n_1=1.34$ and $n_2=2.5-2.8$ correspondingly) was fabricated by standard optical technology using layer by layer deposition of the materials on a glass substrate. The multilayer stack was terminated at both ends with ZnSe layer. The thickness of each layer was  $d_1=d_2=90$~nm. An omnidirectional total reflection was expected within the spectral band $\Delta \lambda=604.3-638.4$~nm with relative bandwidth of 5.3\%. 

Transmission spectra for TE- and TM- polarizations at different incident angles in the range of $0-60^\circ$ were measured using a 'Cary 500'  spectrophotometer (right panel in the figure~\ref{Transmission}). The calculated transmission spectra are depicted in the left panel of the figure~\ref{Transmission}, a good agreement is obtained. From figure~\ref{Transmission} one can see that for spectral range 600--700~nm the transmission coefficient is very low for both polarizations. The absolute values of transmission for TE-polarization in spectral range 630--700~nm was less than 0.001 within the $\pm 0-60^\circ$ aperture, corresponding to a reflection coefficient of 99.9\%.
To reach higher values of angle of incidence a simple set-up consisting of a He-Ne laser and a CCD detector was used. The intensity of laser beam passed through the sample was detected by CCD camera. Sample was mounted on rotational stage to allow different angles of incidence. With this set-up one can directly determine the transmission coefficient of samples at angles
up to $70^\circ$. For larger angles it is necessary to measure the reflection coefficient of samples. The dependence of the transmission coefficients for TE- and TM-polarized incident radiation of a He-Ne laser at 632.8~nm on angle of incidence is presented in the figure~\ref{HeNe}. For TM-polarization circles depict the directly measured transmission coefficient, and squares depict data obtained from reflection measurements. Mismatch between them can be explained by additional reflection from air--ZnSe,  ZnSe--substrate and  substrate--air interfaces. The solid (dashed) curve in the figure~\ref{HeNe} gives theoretically calculated transmission coefficients for TE-(TM-) polarized light.

As can be seen from the figure~\ref{HeNe} the transmission coefficient of TM-polarized radiation remains below $10^{-3}$ over a wide angular range. Due to the Brewster effect at the air--ZnSe, ZnSe--substrate and substrate--air interfaces at large angles it increases to 0.33 at $80^\circ$ and then decreases again. In contrast, transmission of TE-polarized radiation decreases monotonically with growing angle of incidence being less than $10^{-5}$ for angles larger than $40^\circ$. Transmission coefficients of less than $10^{-5}$ are beyond the capabilities of the experimental set-up used. For this reason, the transmitted signal at more than $60^\circ$ cannot be detected. Because of this, no data points for TE-polarization at these angles are presented in the figure~\ref{HeNe}.

A reflectivity of TM-polarized radiation at large angles can be enhanced in the structures terminated at both ends with low index layer, Na$_3$AlF$_6$, (Fig.~\ref{HeNe}).  In this case, transmission at $80^\circ$ is as small as 0.03, corresponding to a reflection coefficient of 97\%.
An overall reflectivity can be enhanced in structures with larger number of layers.

\section{Conclusions}

In summary, we have demonstrated the possibility to achieve a total omnidirectional reflection with  one-dimensional periodic dielectric structures. The origins of the total omnidirectional reflection have been discussed. Optimization criteria of omnidirectional totally reflecting mirror design have been presented. We have found that for reasonable values of structure parameters ($\delta n>1.5, \delta n_0>1.5$) a relatively large omnidirectional total reflection band ($>5\%$) 
may be obtain, making the fabrication of a perfect all-dielectric thin-film mirror feasible. The possibility of partial spontaneous emission control with one-dimensional periodic structures has been discussed. The experimental demonstration of the mirror has been presented at optical frequencies.

\nocite{*}
\bibliographystyle{IEEE}

\end{document}